# Bayesian modelling of lung function data from multiple-breath washout tests

Robert K Mahar[1,2], John B Carlin[1,2,3], Sarath Ranganathan[2,4,5], Anne-Louise Ponsonby[2,6], Peter Vuillermin[2,6,7], and Damjan Vukcevic[1,8]

1. Data Science, Murdoch Children's Research Institute, Parkville, Victoria, Australia
2. Department of Paediatrics, Faculty of Medicine, Dentistry and Health Services, University of Melbourne, Parkville, Victoria, Australia
3. Melbourne School of Population and Global Health, Faculty of Medicine, Dentistry and Health Services, University of Melbourne, Parkville, Victoria, Australia
4. Infection and Immunity, Murdoch Children's Research Institute, Parkville, Victoria, Australia
5. Department of Respiratory and Sleep Medicine, Royal Children's Hospital, Parkville, Victoria, Australia
6. Population Health, Murdoch Children's Research Institute, Parkville, Victoria, Australia
7. School of Medicine, Faculty of Health, Deakin University, Geelong, Victoria, Australia
8. School of Mathematics and Statistics, Faculty of Science, University of Melbourne, Parkville, Victoria, Australia

**Corresponding author**: Damjan Vukcevic, Data Science, Murdoch Children's Research Institute, Flemington Road, Parkville, 3052. Ph: +61 3 9936 6567. Email: damjan.vukcevic@mcri.edu.au

# Abstract

Paediatric respiratory researchers have widely adopted the multiple-breath washout (MBW) test because it allows assessment of lung function in unsedated infants and is well suited to longitudinal studies of lung development and disease. However, a substantial proportion of MBW tests in infants fail current acceptability criteria. We hypothesised that a model-based approach to analysing the data, in place of traditional simple empirical summaries, would enable more efficient use of these tests. We therefore developed a novel statistical model for infant MBW data and applied it to 1,197 tests from 432 individuals from a large birth cohort study. We focus on Bayesian estimation of the lung clearance index (LCI), the most commonly used summary of lung function from MBW tests. Our results show that the model provides an excellent fit to the data and shed further light on statistical properties of the standard empirical approach. Furthermore, the modelling approach enables LCI to be estimated using tests with different degrees of completeness, something not possible with the standard approach. Our model therefore allows previously unused data to be used rather than discarded, as well as routine use of shorter tests without significant loss of precision. Beyond our specific application, our work illustrates a number of important aspects of Bayesian modelling in practice, such as the importance of hierarchical specifications to account for repeated measurements and the value of model checking via posterior predictive distributions.

# Introduction

Chronic adult diseases, and respiratory diseases in particular, often have their origins in early life, and lung function in the early years is a major determinant of lung function in later life [1]. The human lung undergoes rapid development while *in utero*, continuing up until late infancy and, in some cases, well beyond that. Adverse lung growth and development during gestation and infancy have been associated with lifelong deficits in lung function and respiratory health. In this context, techniques are required that enable accurate and feasible measurement of lung function across the life course, to better monitor disease progression and manage responses to therapeutic interventions [2, 3].

The inert gas multiple-breath washout (MBW) test is increasingly used to measure lung function in individuals with chronic and serious respiratory diseases such as cystic fibrosis, as well as in cohort studies investigating the early life origins of asthma and chronic obstructive pulmonary disease. Recent clinical trials have also adopted the MBW as a primary tool for monitoring patient responses to therapeutic interventions [4]. The test begins with a 'wash-in' phase: a mouthpiece is fitted to the patient and, after several breaths of room air, the patient starts inhaling an inert tracer gas mixture until the concentration of inert gas is in equilibrium. At this point, the patient is switched back to room air and the 'washout' phase of the test commences. The molar mass of the respired gases and airflow are measured at high frequency throughout the washout until the lungs are effectively clear of the tracer gas. Specialised software then performs breath detection and derives various quantities that are used to assess lung function. Crucially, the washout phase of the test must be performed during regular, uninterrupted breathing.

In young children the use of MBW is favoured over more conventional lung function tests such as spirometry, since it requires minimal patient co-operation and co-ordination [5, 6]. While lung function can be assessed by a number of different summaries of the MBW test, the most commonly used is the lung clearance index (LCI) [7] which is based on the functional residual capacity (FRC), a widely used measure of lung function in its own right. The widespread adoption of the MBW has culminated in a recent consensus statement that recommends guidelines for MBW analysis and highlights areas that need further research [5].

The data captured during an MBW test and used to calculate the LCI suggest an underlying functional form, yet to our knowledge no one has thus far used an explicit statistical model to exploit this. The standard method for calculating LCI instead uses a simple algorithm based on the observed values [5]. By not modelling the underlying process, the standard method does not in general fully exploit the available information. We propose a relatively simple stochastic model for MBW data that leads to a model-based definition of the LCI. Our approach explicitly models the quantities used to calculate LCI, namely the cumulative expired volume (CEV) and FRC. Based on these explicit definitions of the estimands for LCI, CEV, and FRC, we develop a Bayesian estimation method.

One immediate and novel application of this model is to estimate LCI and FRC from incomplete or shortened MBW tests. Shortening the test procedure is of practical interest because it is time-consuming and prone to disruption, particularly in patients with advanced disease, in pre-school children who have shorter attention spans, and among infants who are being tested during natural sleep. In some cases, breathing irregularities during the wash-in, such as sighs, can be acceptable [8]. However, if breathing irregularities occur during the washout or the patient becomes unsettled, the test data are typically not used and additional testing must be performed or testing abandoned. Indeed, in most studies conducted among sleeping babies, successful MBW measures have only been obtained in 60-70% of tested participants. A number of recent studies have shown that, under certain circumstances, MBW tests can be substantially shortened while still providing clinically meaningful results [9–11]. These studies represent important innovations with the MBW technique and cast doubt on whether the test must be complete before a meaningful summary statistic can be derived. However, they do this by defining a different type of LCI which is based only on earlier time points in the test. In contrast, our approach is to model the available shortened test data, allowing us to meaningfully extrapolate it to directly estimate the usual LCI.

Using data from a large birth cohort study, we assess how our model-based method of estimating LCI, CEV, and FRC compares to the standard method using complete MBW tests. We then assess the performance of our model-based method with shortened MBW tests.

We adopt a Bayesian modelling approach as it allows us to naturally incorporate prior information and account for uncertainty in estimation of our model parameters. By assigning prior distributions to model parameters we are also able to constrain them to more realistic values, which we show is particularly helpful when analysing incomplete tests. We use a subset of tests taken from normal infants to estimate an appropriate informative prior and validate its use on a complementary dataset.

A number of the practical statistical considerations that went into this analysis warrant further explanation. We therefore elaborate on these and consider their broader relevance.

# Modelling MBW data

## Data description

The Barwon Infant Study (BIS) is a birth cohort study in the Barwon region of south-eastern Australia designed to investigate the impact of early-life exposures on immune, allergic, cardiovascular, respiratory, and neurodevelopmental outcomes (at inception $n = 1,074$) [12].

MBW data were collected from infants aged between 4 and 12 weeks between February 2011 and December 2013. Acceptable tests exhibited a stable breathing pattern with no artefacts such as sighs, sucking, snoring, or mask leaks, or any observable software artefacts. We obtained a total of 1,197 acceptable MBW tests from 432 infants in this study. Of these infants, 89 (21%) had only a single test performed, while 115 (27%), 98 (23%), 81 (19%), 35 (8%), 13 (3%), and 1 (<1%) participants had 2, 3, 4, 5, 6, and 7 replicate measurements available, respectively. MBW tests were performed in accordance with current guidelines [5]. Infants were tested during natural sleep. All tests were performed with 4% sulphur hexafluoride ($SF_6$) using a mainstream ultrasonic flowmeter (Exhalyzer D, Ecomedics, Duernten, Switzerland). For the purposes of this analysis, MBW data were processed with WBreath (version 3.19, ndd Medizintechnik AG, Zurich, Switzerland). There is limited native support for data processing and export in WBreath [13], so where possible we automated these functions via the graphical user interface using AutoIt, a scripting language [14].

## MBW model

The data collected as part of a routine MBW test consist of air flow (L/s) and molar mass of the gas mixture (g/mol) measured over a short period of time (approx. 1–3 min) at high frequency (200 Hz); see Figure 1. The wash-in component of the test is not typically of interest and is thus discarded. From the subsequent washout phase of the test, the following quantities are derived (Figure 2): the tracer gas quantity (GAS; the current amount of tracer gas, as a proportion of the amount of tracer gas at the start of the washout), cumulative expired volume of gas mixture (CEVGM), and the cumulative expired volume of tracer gas (CEVTG). As is typically done, we derive discrete versions of these quantities by calculating them on a per-breath basis (the quantities are therefore interpretable as representing the state at the end of each breath cycle) [5, 15].

We first introduce notation for modelling the data from a single MBW test. Let $k = \{0, 1, \ldots, K\}$ index breaths from the start of the washout phase (indexed by $k = 0$). We denote the GAS, CEVGM, and CEVTG series as $c(k)$, $v(k)$, and $r(k)$ respectively. Figure 2 shows $c(k)$, $v(k)$, and $r(k)$ for 10 randomly selected MBW tests. The LCI as defined in the aforementioned consensus statement, herein referred to as the 'standard method', is denoted here as

$$LCI^{(s)} = \frac{CEV^{(s)}}{FRC^{(s)}} = \frac{v(k^{(40)})}{r(k^{(40)})/\left(1 - c(k^{(40)})\right)}, \qquad (1)$$

where the cumulative expired volume ($CEV^{(s)}$) denotes the CEVGM measured at breath $k^{(40)}$, the functional residual capacity ($FRC^{(s)}$) denotes an estimate of the volume of a typical expiration (using quantities measured at breath $k^{(40)}$), and the parenthetical $s$ denotes the 'standard method'. As recommended in the consensus guidelines [5], the $k^{(40)}$ used to calculate the quantity in equation (1) is the first of three consecutive breaths for which $c(k) \leq 1/40$. We define a complete test to be one that allows identification of $k^{(40)}$; i.e. it must include data measured up to the third consecutive breath for which $c(k)$ is under $1/40$.

Our approach was to specify statistical models for $c(k)$, $v(k)$, and $r(k)$, with each quantity represented by a smooth function plus an independent random error process.

We introduce notation for vectors of parameters, $\boldsymbol{\beta} = \{\beta_1, \beta_2, \ldots, \beta_5\}$ and $\boldsymbol{\sigma} = \{\sigma_c, \sigma_v, \sigma_r\}$, which will be used to characterize models for the three quantities of interest. We modelled the tracer gas quantity $c(k)$ using a two-phase exponential decay function of $k$ for the (log-scale) mean, which is implied by a theoretical two-compartment lung model [16]:

$$f(k; \boldsymbol{\beta}) = \beta_0 \exp(-\beta_1 k) + (1 - \beta_0) \exp(-\beta_2 k), \qquad (2)$$

where $0 < \beta_0 < 1$ is the proportionate weight of the fast decaying component and $\beta_1$ & $\beta_2$ ($0 < \beta_1 < \beta_2$) denote the fast and slow decay constants respectively. This model is commonly applied to various physical systems, including the lungs, and at least one recent study has applied the model directly to MBW end-tidal tracer gas concentration in adults, albeit to achieve different aims [17]. Since the data are restricted to be positive, we model $c$ as log-normal with independent and identically distributed errors:

$$c(k) \sim \log N(f(k; \boldsymbol{\beta}), \sigma_c^2). \tag{3}$$

The cumulative expired volume of gas mixture, $v(k)$, was modelled as an increasing linear function of $k$, with geometric mean

$$g(k; \boldsymbol{\beta}) = \beta_5 k, \tag{4}$$

and the cumulative expired volume of tracer gas, $r(k)$, as an increasing exponential decay function of $k$, with geometric mean

$$h(k; \boldsymbol{\beta}) = \beta_3 (1 - \exp(-\beta_4 k)). \tag{5}$$

Note that the functions $v$ and $r$ are both strictly positive and monotonically increasing, which we capture by modelling their discrete derivatives (denoted here with primes), and do so on a log scale:

$$\ln(g'(k; \boldsymbol{\beta})) = \ln(\beta_5), \tag{6}$$

$$\ln(h'(k; \boldsymbol{\beta})) = \ln(\beta_3) + \ln(1 - \exp(-\beta_4)) - \beta_4 k. \tag{7}$$

An added benefit of these transformations is that we can assume independent and identically distributed errors on a natural scale for each quantity, giving us the final model:

$$v'(k) \sim \log N(g'(k; \boldsymbol{\beta}), \sigma_v^2), \tag{8}$$

$$r'(k) \sim \log N(h'(k; \boldsymbol{\beta}), \sigma_r^2). \tag{9}$$

Using this model, we define the end-test breath, which we denote $\theta$, as the real-valued breath index at which the expected value of $c$ is equal to $1/40$:

$$f(k = \theta; \boldsymbol{\beta}) = 1/40. \tag{10}$$

We can thus redefine FRC, CEV, and therefore LCI, in terms of the model parameters:

$$CEV^{(m)} = g(\theta; \boldsymbol{\beta}), \tag{11}$$

$$FRC^{(m)} = \frac{h(\theta; \boldsymbol{\beta})}{1 - f(\theta; \boldsymbol{\beta})}, \tag{12}$$

$$LCI^{(m)} = \frac{g(\theta; \boldsymbol{\beta})}{h(\theta; \boldsymbol{\beta})/(1 - f(\theta; \boldsymbol{\beta}))}, \tag{13}$$

where the parenthetic $m$ denotes the 'model-based' method. Note, however, that the standard method defines FRC in terms of a simple summary statistic, by taking the observed cumulative expired volume of tracer gas and dividing it by $1 - c(k)$ at breath $k^{(40)}$, where some gas will remain in the lungs. Using the model-based method a more natural approach is to use the asymptotic value of $FRC^{(m)}$, which represents the state of the lungs when completely free of the inert gas:

$$\lim_{\theta \to \infty} \frac{h(\theta; \boldsymbol{\beta})}{(1 - f(\theta; \boldsymbol{\beta}))} = \beta_3. \tag{14}$$

Thus we propose a more refined model-based definition, denoted by an asterisk, with $FRC^{(m^*)} = \beta_3$, giving:

$$LCI^{(m^*)} = \frac{g(\theta; \boldsymbol{\beta})}{\beta_3}. \tag{15}$$

Given an end-test threshold 1/40 (n.b. the numerator still depends on this threshold), the $LCI^{(m^*)}$ is therefore wholly defined as a function of the model parameters $\boldsymbol{\beta}$.

We fitted the models using a Bayesian approach, assigning either diffuse or informative prior distributions to the parameters.Stan, a probabilistic programming language [18], was used for parameter estimation (see Appendix). Diffuse distributions contained only vague information based on the potential scale and theoretical support of a parameter [19]. Therefore we assumed that $\beta_0 \sim \text{Beta}(2,2)$, $\beta_1, \beta_2, \beta_4 \sim \text{Normal}(0,1)$, $\beta_3 \sim N(0, 10^3)$, and $\beta_5 \sim N(0, 10^2)$. These priors have the advantage over wider, less informative priors in that they aid computation and inference while providing only minimal information. We used half-Cauchy distributions as diffuse priors for the standard deviations, as recommended [19, 20], such that $\boldsymbol{\sigma} \sim \text{Half-Cauchy}(0, 2.5)$. Informative prior distributions, on the other hand, were intended to characterise the typical distribution of the MBW model parameters and took the form of a multivariate normal distribution such that $\boldsymbol{\beta} \sim \text{MVN}(\boldsymbol{\mu}, \boldsymbol{\Sigma})$, where $\boldsymbol{\mu}$ denotes the mean vector and $\boldsymbol{\Sigma}$ denotes the variance-covariance matrix. The informative priors were obtained by fitting a two-level hierarchical model for GAS, CEVGM, and CEVTG to a subset of our data, as outlined in the next section.

We fitted the models with both diffuse and informative priors to individual tests with complete data and calculated the posterior medians of $CEV^{(m)}$, $FRC^{(m^*)}$, and $LCI^{(m^*)}$ in both cases, which we refer to herein as the 'model-based' estimates.

## Obtaining an informative prior via a hierarchical model

The parameters of the informative prior distributions used in this paper were estimated using hyperparameters obtained by fitting a two-level hierarchical model for GAS, CEVGM, and CEVTG. This model allows the sharing of information between tests by modelling the lower-level per-test parameters (or random effects) as coming from a common distribution defined by the hyperparameters. This allowed us to summarise and extract the relevant information from a large set of tests into a form suitable for use as a prior distribution. Because our primary interest here was to approximate prior distributions for use with individually modelled tests, we deliberately used a simple two-level hierarchical model rather than a more complex three-level model that would account for the fact that we had multiple replicates for each participant. To ensure that we were using data appropriate to this simpler model specification, we reduced the data by taking only the first replicate test for each patient, resulting in 414 tests. We adopted a simple holdout/validation approach whereby we fitted this two-level model to one half, or 212, of the tests in this dataset. The posterior medians of the hyperparameter distributions were extracted for use as parameters of the informative prior distributions on the validation data.

We define hyperparameters $\boldsymbol{\beta}$ with lower-level parameters $\boldsymbol{w_i} = \{w_{1i}, w_{2i}, \dots, w_{6i}\}$, a combined vector of random effects for test $i = \{1, \dots, 212\}$. We specify our hierarchical model as follows. First, the underlying functional forms for each series were:

$$f(k; \boldsymbol{\beta}, \boldsymbol{w_i}) = (\beta_0 + w_{0i}) \exp((-\beta_1 + w_{1i})k) + (1 - (\beta_0 + w_{0i})) \exp((-\beta_2 + w_{2i})k)$$

$$g(k; \boldsymbol{\beta}, \boldsymbol{w_i}) = (\beta_5 + w_{5i})k$$

$$h(k; \boldsymbol{\beta}, \boldsymbol{w_i}) = (\beta_3 + w_{3i})(1 - \exp((-\beta_4 + w_{4i})k))$$

The distribution of our observations for a test $i$ can now be summarised as follows:

$$c_i(k) \sim \text{logN}(f(k; \boldsymbol{\beta}, \boldsymbol{w_i}), \sigma_c^2)$$

$$v_i'(k) \sim \text{logN}(g'(k; \boldsymbol{\beta}, \boldsymbol{w_i}), \sigma_v^2)$$

$$r_i'(k) \sim \text{logN}(h'(k; \boldsymbol{\beta}, \boldsymbol{w_i}), \sigma_r^2)$$

We specify a multivariate normal distribution for the random effects: $\boldsymbol{w_i} \sim \text{MVN}(0, \boldsymbol{\Sigma})$, where $\boldsymbol{\Sigma}$ is the variance-covariance matrix of the random effects $\boldsymbol{w_i}$ and has a correlation matrix $\boldsymbol{P}$. As is common practice in Stan, we define a prior on the correlation matrix, $\boldsymbol{L} \sim \text{LKJ}(2.0)$, where $\boldsymbol{L}$ is the Cholesky decomposition of $\boldsymbol{P}$ and $\boldsymbol{w_i} = \text{diag}(\boldsymbol{\sigma_w})\boldsymbol{LZ}$ where $\boldsymbol{Z}$ is a matrix of unit normal random variables corresponding to the individual random effects, this prior and parameterisation implies that the off-diagonals of the correlation matrix are near zero and is consistent with a multivariate non-centred parameterisation on the random effects $\boldsymbol{w_i}$ [19, 21]. All hyperparameters contained in $\boldsymbol{\beta}$ were specified with diffuse priors as outlined in the previous section. Note that we assume that the standard deviations are fixed across tests and $\boldsymbol{\sigma} \sim \text{Half-Cauchy}(0, 2.5)$.

## Results

We fitted the hierarchical MBW model to 209 of the 212 MBW tests in the holdout subset. Three of the tests were excluded as the data appeared highly irregular. Parameter convergence was sufficient, with $\hat{R}$ values close to 1, and effective sample sizes ranged across the hyperparameters from 102 to 4,000 (median: 452). A summary of the posterior distributions used to create informative priors is shown in Table 1, along with the relevant correlation matrix.

We also fitted the individual MBW model using diffuse priors to each of the 1,197 individual tests using diffuse priors, and to each of 212 individual tests in the validation subset using the informative priors. Parameter convergence was sufficient in most cases, with $\hat{R}$ values close to 1. Although for a small number of tests some parameters exhibited $\hat{R}$ >1.1, we judged that this was unlikely to materially affect the overall results. Effective sample sizes across all tests were judged as sufficient with a median of 1,141 (IQR: 841, 1,378) and 1,207 (IQR: 900, 1,418) for the diffuse and informative priors respectively.

The model fit for a typical individual's MBW data with diffuse prior information is shown in Figure 3, along with 10 simulated datasets drawn from the posterior predictive distribution for each quantity. The model appears to describe the data well on the transformed scale (left column), and on the original scale (right column) (for detailed description of model fit, see *Further considerations*). Although we only show a single test in Figure 3, we observed similar results for other tests.

Upon fitting the above models, we examined the residual errors for the presence of autocorrelation via visual inspection of the residual partial autocorrelation function of 100 randomly selected tests (data not shown). Additionally we fitted the models described in equations (3), (8), and (9) to the same 100 individual tests, modifying the error term to have an auto-regressive structure with lag 1 (AR1). Overall, there was insufficient evidence to suggest the presence of a clear autocorrelation structure or differences in the estimates of MBW outcomes when using either independent or AR1 errors. We therefore retained the assumption of independent errors for each model.

## Assessing agreement and variance

### Model

We assessed agreement between the standard and model-based methods using a Bayesian variance components model. Since the MBW tests in our study were performed multiple times on the same participant, we needed a model flexible enough to account for replicate measurements. We used a variation of a model proposed by Carstensen et al. [22], which accounts for variation between participants and also variation due to replicate measurements. Whereas Carstensen et al. [22] model the between-participant variation using fixed effects, in the Bayesian setting, and with a large number of participants, we found that these were more naturally modelled as random effects.

We used the above model solely for comparing the standard method and the model-based method with diffuse prior, in order to examine the agreement between the methods without influence from informative prior information. Using the additional information will simply tend to shrink the model-based MBW outcomes toward the prior mean, so we judged that a comparison with the informative prior was unnecessary. Unlike standard agreement analysis, our goal here is not to establish strict equivalence or interchangeability between the two methods, but rather to examine how they differ and whether we are satisfied that any differences are justified. The use of a proper agreement model for this purpose allows us to properly control for the replicate structure within our data.

Denoting the method of measurement by $m$, where $m = 1$ denotes the standard method, $m = 2$ denotes the model-based method using diffuse priors, the study participants by $p$ ($p = 1, \ldots, 432$), and the $r$'th replicate of method $m$ on participant $p$ as $y_{mpr}$, the agreement model was defined as:

$$y_{mpr} = \alpha_m + u_p + a_{pr} + c_{mp} + e_{mpr},$$

$$u_p \sim \text{N}(0, \gamma), \quad a_{pr} \sim \text{N}(0, \omega), \quad c_{mp} \sim \text{N}(0, \tau_m), \quad e_{mpr} \sim \text{N}(0, \sigma_m), \qquad (16)$$

$$\alpha_m \sim \text{N}(0, 10^4), \quad \gamma, \omega, \tau_m, \sigma_m \sim \text{Half-Cauchy}(0, 2.5),$$

where $\alpha_m$ is the mean effect for method $m$, the variation between participants is captured by $\gamma$, the variation between replicates (nested within participants) is captured by $\omega$, the extra variation between participants attributable to each method is captured by $\tau_m$, and the within-participant variation for a specific method $m$ (i.e. across replicates) is captured by the residual standard deviation $\sigma_m$. Note that the inclusion of the $a_{pr}$ terms makes this correspond to the 'linked replicates' (referred to as 'non-exchangeable replicates' henceforth) model of Carstensen et al. [22]. As we are comparing only two methods, the different $\tau$ parameters become indistinguishable, so we set $\tau_1 = \tau_2 = \tau$; furthermore, the size of this component was estimated to be negligible in our data.

We encountered numerical issues fitting the full model that could not be resolved through either model reparameterisation or fine tuning of the sampler parameters. Therefore, although our data follow a non-exchangeable structure (since the input data used by each method are identical, meaning that the output estimates correspond to each other on a per-replicate basis), we assumed exchangeable replicates, dropping the $\alpha_{pr}$ terms (which have the same value across methods, and thus capture the 'linking' structure). This allowed us to estimate a separate residual standard deviation for each method, albeit with the variation due to replicates, which would otherwise be captured by $\omega$, being distributed among the estimable variance components $\gamma$, $\tau$ and $\sigma_m$. However, it was also of interest to understand the size of the per-replicate variation, so we fitted a second model where we included $\alpha_{pr}$ but assumed a common residual standard deviation for each method; i.e. $\sigma_1 = \sigma_2 = \sigma$.

We again took a Bayesian modelling approach, using Stan for estimation (see Appendix), and assigned diffuse priors on the model parameters [19, 20]: normal distributions for the mean components, and half-Cauchy distributions for the variance components, as shown in (16).

## Results

The two variance components models were fitted to the 1,197 MBW tests using Stan. Parameter convergence was sufficient in all cases, with $\hat{R}$ values close to 1. Effective sample sizes were also judged as sufficient, ranging from 70 to 4,000 (median: 2,929) and 977 to 4,000 (median: 483) across all mean and variance components for the exchangeable replicates model and the non-exchangeable replicates model, respectively.

Comparisons of LCI, CEV, and FRC estimated by each method, relative to the standard method, are shown in Figure 4 (in this figure we also show, for reference only, a comparison with the estimates when using an informative prior, but as explained above we do not discuss these in this section). Estimates of the parameters of the variance components model assuming exchangeable replicates and method-specific residuals are summarised in Table 2.

For CEV and FRC, the model-based estimates were slightly smaller on average (looking at the differences between mean components). This was presumably due to the use of the model-based end-test definition which, by nature, is typically earlier than when using the standard method. Nevertheless, the resulting estimates of LCI appeared to be similar between both methods.

The residual variance component of the model-based LCI was slightly higher than under the standard method. In contrast, for CEV it was the other way around, which may be explained by the different end-test definitions (see previous paragraph). However, all of these differences are minor when compared to the overall magnitude of variation.

The method-participant interaction parameter, $\tau$, was relatively small for each outcome, which was not surprising since we were comparing similar methods. Interestingly, for CEV and FRC the within-participant (residual) variation for each method was about half the size of the between-participant variation $\gamma$, but for LCI these were roughly similar to each other.

**Error! Reference source not found.** shows estimates from the variance component model assuming non-exchangeable replicates and a single common residual term. The main purpose of fitting this model was to investigate the extent to which the exchangeability assumption affected our interpretation of the previous model. As the table shows, the mean components remained similar, but a substantial amount of variation was 'redirected' to the replicate-participant interaction term $\omega$ from the between-participant interaction $\gamma$ and the residual $\sigma$. This suggests that a substantial amount of the residual variation summarised in Table 2 is in fact due to the variation between replicates, meaning that the scope for substantial differences between methods in this regard is actually rather limited.

Taken together, these results suggest that the diffuse-prior model produced estimates of LCI that could be used interchangeably with the standard method. Furthermore, they suggest that the standard method is in fact a good estimator of MBW outcomes using complete data.

## Assessing shortened tests

### Method

We assessed the accuracy of our model when using shortened test data. For each test, we truncated each of $c(k)$, $v(k)$, and $r(k)$ up to the first $k$ where $c(k)$ fell below some threshold and remained there for two subsequent breaths, in the same way that the standard method is used to empirically determine the end-test breath. This method of truncation is similar to that used by other studies that have evaluated the use of 'shortened' MBW tests [9–11]. We used six different thresholds for the tracer gas quantity: 1/3, 1/4, 1/5, 1/10, 1/20, and 1/30. We fitted the MBW models to the artificially shortened test data to estimate, by posterior medians, the model-based outcomes LCI and FRC using either the diffuse or informative prior.

We assessed the accuracy of our model, for each of the six truncation thresholds, by summarising estimates of LCI and FRC from shortened tests relative to estimates obtained using complete test data (to make them comparable across tests). We summarised these by examining the distribution of the relative prediction errors and the width of the 95% credible intervals (CI) of the predictions.

## Results

We fitted the MBW models to the six artificially shortened versions of all 1,197 MBW tests using diffuse priors, and similarly also to the 212 MBW tests in the validation sample using informative priors.

In total, this resulted in close to 8,500 separate analyses. Since it is impractical to assess the convergence of so many tests graphically, for these analyses convergence was assessed to be sufficient for a single test if all parameters had scale reduction factors ($\hat{R}$) below 1.25. In general, as shown in Table 4, relatively few tests exhibited sub-optimal MCMC convergence, with the exception of the scenarios that had very little data. In any case, where convergence was not sufficient, it is likely that finer tuning of the MCMC sampler or additional sampling iterations would have resolved the issue. Regardless, dropping the estimates from tests that had convergence issues had negligible impact on these results for any of the diffuse and informative prior information cases, and thus we did not remove them from the results shown here.

Figure 5 shows plots of the median and the 2.5th and 97.5th percentiles of the prediction errors and the width of the 95% credible intervals, for each of the shortened datasets across either 1,197 MBW tests (in the diffuse prior case) or 212 MBW tests (in the informative prior case), as a percentage of the corresponding model-based LCI estimate using complete data. Figure 6 shows a similar summary for estimates of FRC.

For LCI, in both the diffuse and informative prior information cases, the accuracy of the model-based estimate clearly improved with more data. Furthermore, in each shortened test scenario, the median prediction error was less than 10% under either prior. This indicates that the model-based estimates are generally accurate even when the data are highly incomplete. The informative priors helped to moderate the estimates, particularly when the tests were highly incomplete, somewhat reducing the bias that was discernible with the diffuse priors. The range of the prediction errors shrank as we considered longer tests, showing how the estimates become progressively more reliable with additional data. For shortened test thresholds higher than 1/10, the model was clearly more reliable when using informative priors.

As shown in Figure 6 the results were similar for estimation of FRC, although the model-based FRC showed little to no discernible bias in either the diffuse and informative prior cases, and the range of the prediction errors was smaller than that for LCI in both cases, with the exception of tests shortened to tracer gas quantities higher than 1/10 in the diffuse case.

The strong downward bias of LCI for very short tests when using diffuse priors can be understood in the context of the underlying models. The log discrete derivative models, used for CEVGM and CEVTG, are linear and thus can be extrapolated with some accuracy beyond their range. In contrast, the underlying model for $\theta$ is a mixture of fast and slow decaying exponential components. With only a limited amount of data available (due to short tests) and the absence of informative priors to constrain the two components, the model placed a greater weight on the fast decaying component, which dominates in the earlier stages of the test, resulting in a smaller $\theta$, and thus smaller $CEV^{(m)}$ and $LCI^{(m^*)}$. This was somewhat rectified by using informative priors, which ensured that the model properly acknowledged the existence of a slowly emptying compartment even when there were insufficient data to easily estimate it, resulting in a more accurate estimate of $\theta$, and correspondingly better estimates of $CEV^{(m)}$ and $LCI^{(m^*)}$.

A clear issue with both the LCI and FRC estimates using shortened tests was the degree of uncertainty around each estimate. This is assessed in both Figure 5 and Figure 6 by summarising the proportional width of the 95% CI across the estimates for each shortened test scenario. Uncertainty around estimates of both $FRC^{(m^*)}$ and $LCI^{(m^*)}$ using data shortened to thresholds higher than 1/10 is high enough to discourage any use of estimates from tests shorter than this level. However, the uncertainty appeared to stabilise at a level similar to the complete data case at thresholds equal to or lower than 1/10, particularly when using informative prior information. The fact that the uncertainty did not reduce beyond this threshold is likely representative of the inherent uncertainty in the underlying LCI definition. Moreover, this suggests that there is little to be gained by performing the test for any longer than up to a tracer gas quantity threshold of around 1/10.

Overall, we were able to obtain reasonable estimates of LCI and FRC even with data shortened to a tracer gas quantity threshold as high as 1/10. On average, a complete test is 30 breaths in length, and a test shortened to the 1/10 tracer gas level is approximately 17 breaths, translating to an average 43% reduction in breaths required. For tests shortened to thresholds higher than this, our estimates became inaccurate. This suggests some portion of the test beyond the start but before the 1/10 threshold contains information necessary to determine LCI at the end of the test. This makes intuitive sense: because the areas of the lung with least ventilation are likely to take longer to empty, the slope of the tracer gas quantity curve is flatter later in the test.

Predictions made without any data from this part of the test will typically be underestimates because $\theta$ will typically be underestimated. From these results, it appears that, in general, sufficient additional information to reliably estimate LCI becomes available between proportions 1/5 and 1/10 (on average this gap spans approximately five breaths).

## Further considerations

### Modelling cumulative data

The wide variation of the posterior predictive distribution of the CEVTG model on the cumulative scale, as shown in Figure 7, may appear at odds with the observed data. Note, however, that a key feature of the data was the strict monotonicity of the CEVGM and CEVTG series.

We initially specified our model with independent errors directly on the cumulative scale. While this produced posterior median estimates that fit the observed data closely, it also gave rise to posterior predictive distributions that did not reflect the nature of the observed data. In particular, the model violated the monotonicity constraint for both series (see Figure 7). We also found that the posterior distributions of the parameters did not reflect the observed variation across replicates, with credible intervals for FRC (which is derived from CEVTG) being overly narrow (see Figure 8).

In thinking about alternative models, shifting to the discrete derivative on a log scale seemed a natural choice in order to preserve monotonicity. When viewed on the cumulative scale, this resulted in series with seemingly overly wide variation. However, upon reflection we realised that such variation is actually probably indicative of reality and reflects the nature of these data. Namely, CEVTG is a cumulative measure of a decreasing quantity, meaning that the observed CEVTG curves are heavily influenced by any noise in the early breaths of the washout. Therefore the observed asymptote may actually be inaccurate and models should not necessarily try to fit it too closely. Indeed, we found that the posterior variation given by our discrete derivative model matched quite well with the actual variation we observed between replicate tests (see Figure 8).

### Hyperparameter estimation to obtain informative priors

Our analysis of shortened tests showed the utility of using a parametric model along with genuine prior information with this type of extrapolation problem. Additionally, the complete data analysis of CEVTG curves suggests that using informative prior distributions also shrinks the parameter estimates towards more realistic values. More generally, there is no doubt opportunity to routinely use prior information to aid the analysis of individual 'growth/decay-like' data and to estimate unseen quantities of these models where data are limited. This approach should be particularly useful when it is costly or difficult to collect additional data for an individual and one has a solid model of the underlying 'growth' or 'decay' pattern, as in the case of the MBW test.

We believe our approach to obtaining informative priors for the MBW model—using the hyperparameters of a hierarchical model—is appropriate and effective for this problem. It builds on the structure of the simpler individual-level model in the right way to allow us to pool information across a dataset. We were able to validate its effectiveness through a (holdout) validation step, to ensure we were not 'using the data twice'. Other researchers are now able to use our informative prior directly.

**Variance components model**

We took the approach of examining the properties of our estimation procedures by comparing our model-based results with the traditional empirical approach using a dataset of several hundred healthy children and a variance components model. This was in an applied Bayesian spirit in contrast to a frequentist evaluation of repeated-sampling properties via simulation. This approach also proved useful as it allowed us to better understand the source of the relatively high observed variance in MBW outcomes (which is informative for future analyses with a more epidemiological focus).

When analysing the residual variances of each method we made the assumption of an exchangeable replicate structure. While it would have been useful and convenient to parse out the variation into individual components to the fullest extent possible, by including an additional interaction term to account for the non-exchangeability of our replicates, we found that it was not possible to do so while achieving sufficient mixing of the posterior chains even after applying common reparameterisations and modifying the operating parameters of the sampler.

Faced with this type of situation, the modeller has one of three practical choices. They could drop an interaction term and assume exchangeability, allowing estimation of separate residual variance components, with the interpretation that those components contain some variation due to replicates within each participant. They could include retain the interaction term for non-exchangeable replicates but assume a common residual variance. Or, as we have done, do both and compare the results.

# Conclusions

As far as we know we are the first to use statistical models to define and then estimate the key parameters of the MBW, in particular LCI and FRC. By modelling the MBW data rather than using simple summaries, we developed a principled way of assessing MBW outcomes. These models are strongly tied to the underlying physiology, and therefore may reveal patient characteristics that are unobservable using simpler methods. Additionally, we applied these models to assess how much test data are required to obtain reasonable estimates of MBW outcomes and how the uncertainty around these estimates changes given different degrees of test completion.

Applying our model to complete MBW tests, we observed posterior predictive distributions that reflected the underlying structure and variation in the observed data. As part of this modelling process we also examined the variation structure of the MBW parameters in depth using a variance components model. Our complete-data model-based estimates of LCI have good agreement and similar residual variance compared to the same quantities estimated using the standard empirical method. This suggests that the standard method is actually a good estimator of LCI using complete data, something that would have been difficult to establish without implementing a model-based approach. Using informative priors seems to constrain the estimation procedure to produce more realistic values of FRC when faced with noisy data. Our exposition of a model-based LCI predicted from shortened tests provides statistical support to recent studies that have found that an LCI defined using a higher tracer gas quantity threshold can still be clinically meaningful, e.g. $k^{(20)}$ and in some cases $k^{(10)}$, instead of $k^{(40)}$ [9–11].

Our model can produce reliable estimates of LCI from MBW tests that are, on average, 43% shorter than a complete test (measured by number of breaths). This potentially allows data that would otherwise be unused, due to artefacts or test disruption, to be shortened and subsequently analysed to estimate key quantities of lung function with good accuracy. Given the logistical challenges of conducting MBW testing in unsedated infants during natural sleep, this is an important finding as it may allow the various cohort studies that have undertaken MBW testing to utilise a larger proportion of their MBW data. At a clinical level, this new approach may save both patient and clinician time and effort as potentially fewer tests will be required to achieve the desired number of acceptable LCI measurements. The analyses reported here were limited to a population-derived infant sample. It is not clear how our model would perform with MBW data from participants with lung disease, but it is likely that for these participants there will be an even more pronounced change in the slope in the later part of the tracer gas quantity curve, which could potentially make extrapolation into this area of the test less accurate using only shortened test data with diffuse prior information. A clear direction for future research is therefore to assess these models using MBW tests from both healthy and disease-affected infants.

We note a number of more general aspects of our analyses. In particular, the need to model underlying increments of cumulative data, particularly where there are monotonicity constraints, the utility of informative priors in 'growth' or 'decay' type problems where a realistic data model exists, and difficulties fitting a Bayesian variance components model with non-exchangeable replicates.

# Acknowledgments

The authors are grateful to the BIS study participants and their families for their contribution, and to Louise King and Hilianty Tardjono for their assistance in obtaining and processing the MBW data. We also thank the BIS investigator team for their collaboration in making this work possible.

This work was supported by the National Health and Medical Research Council: Centre of Research Excellence Grant ID 1035261 (Victorian Centre for Biostatistics; ViCBiostat), Senior Research Fellowship ID 1110200 (Anne-Louise Ponsonby), and Project Grant ID 1009044. Robert Mahar was supported by the Australian Government Research Training Program Scholarship. The MBW testing equipment was purchased by the Shane O'Brien Memorial Asthma Foundation. Research at the Murdoch Children's Research Institute is supported by the Victorian Government's Operational Infrastructure Support Program.

# Appendix

## Model estimation

We took a Bayesian approach to estimating our model parameters for a number of reasons. First, the approach makes it relatively straightforward to define and estimate derived parameters such as the model-based LCI by sampling from the posterior distribution using Stan, a probabilistic programming language [18]. This allows us to quantify the uncertainty around our model-based LCI using Markov chain Monte Carlo (MCMC) methods. Second, the Bayesian approach allows us to naturally incorporate prior information in our model parameters. By assigning prior distributions to our model parameters we are able to constrain them to realistic values which can be effectively updated with the available data given different levels of test completion. The Stan model code and a description of the data used are provided in the Supplementary material.

We sampled from the posterior distribution of each parameter estimated in this paper using the Hamiltonian Monte Carlo (HMC) algorithm implemented within Stan, called from the R software environment [23]. In each case, four MCMC chains were run in parallel comprising a discarded 'warm-up' phase and a 'sampling' phase of 1,000 iterations each with the latter, comprising 4,000 samples from the posterior distribution being used for parameter estimation.

In line with Gelman et al. [24], we judged a sufficient effective sample size to be $5m$, where $m = 2 \times$ number of chains. Thus with 4 chains we aimed to have all parameters exhibit an effective sample size of at least 40 to be sufficient for accurate parameter inference. Therefore, where effective sample sizes appeared sub-optimal, we state this in the main text.

Convergence of the Markov chains was assessed both graphically, via visual inspection of the MCMC chains, and statistically, by ensuring that the scale reduction factor $\hat{R}$ was near 1 for each parameter [24].

# Tables

## Table 1

**Summary of the hierarchical model posterior.** Posterior medians and 95% credible intervals of the hyperparameters in the model. 'SD' refers to the standard deviation of the random effects associated with each hyperparameter. For clarity, the correlation matrix shows the correlations multiplied by 100 (e.g. 0.24 is shown as 24).

| Parameter | Posterior summary | | Correlations | | | | |
|---|---|---|---|---|---|---|---|
| | Median | SD | $\beta_1$ | $\beta_2$ | $\beta_3$ | $\beta_4$ | $\beta_5$ |
| $\beta_0$ | 0.68 (0.65, 0.7) | 0.15 (0.14, 0.17) | - | 25 (11,36) | 83 (75,89) | 2 (-12,17) | -15 (-28,-2) |
| $\beta_1$ | 0.129 (0.126, 0.13) | 0.023 (0.021, 0.025) | - | - | 25 (8,40) | -11 (-25,4) | 91 (88,94) |
| $\beta_2$ | 0.53 (0.49, 0.58) | 0.2 (0.16, 0.23) | - | - | - | -8 (-25,8) | -10 (-26,7) |
| $\beta_3$ | 124.3 (121.85, 126.97) | 18.27 (16.5, 20.34) | - | - | - | - | -14 (-27,0) |
| $\beta_4$ | 0.137 (0.134, 0.14) | 0.024 (0.022, 0.026) | - | - | - | - | - |
| $\beta_5$ | 29.96 (29.09, 30.85) | 6.38 (5.83, 7.04) | - | - | - | - | - |

## Table 2

**Agreement analysis: exchangeable replicates, method-specific residuals.** Parameter estimates (posterior medians and 95% credible intervals) from the variance components model fitted to the BIS MBW outcomes, comparing standard and diffuse prior model-based estimates.

|  | LCI | | CEV (mL) | | FRC (mL) | |
|---|---|---|---|---|---|---|
| Parameter | Median | 95% CI | Median | 95% CI | Median | 95% CI |
| ***Mean components*** | | | | | | |
| Standard method ($\alpha_1$) | 6.10 | (6.07, 6.14) | 796.40 | (785.28, 808.21) | 131.10 | (129.13, 133.15) |
| Model-based method: diffuse ($\alpha_2$) | 6.08 | (6.05, 6.12) | 773.28 | (761.96, 785.12) | 127.90 | (125.94, 129.93) |
| *Mean difference ($\alpha_1 - \alpha_2$)* | 0.02 | (-0.01, 0.05) | 23.14 | (19.07, 27.34) | 3.20 | (2.51, 3.94) |
| ***Variation components*** | | | | | | |
| Between-participant ($\gamma$) | 0.31 | (0.28, 0.34) | 110.38 | (103.38, 119.63) | 20.50 | (19.22, 21.96) |
| Method-participant interaction ($\tau$) | 0.01 | (0, 0.03) | 0.84 | (0.03, 2.97) | 0.18 | (0.01, 0.57) |
| Residual components | | | | | | |
| Standard method ($\sigma_1$) | 0.35 | (0.33, 0.36) | 52.58 | (50.4, 54.93) | 8.70 | (8.32, 9.07) |
| Model-based method: diffuse ($\sigma_2$) | 0.39 | (0.37, 0.41) | 48.86 | (46.86, 51.05) | 9.00 | (8.66, 9.41) |
| *Residual ratio ($\sigma_1/\sigma_2$)* | 0.89 | (0.84, 0.95) | 1.08 | (1.02, 1.14) | 1.00 | (0.91, 1.02) |

Notes: LCI: lung clearance index; CEV: cumulative expired volume; FRC: functional residual capacity; CI: credible interval.

## Table 3

**Agreement analysis: non-exchangeable replicates, common residuals.** Parameter estimates (posterior medians and 95% credible intervals) from the variance components model fitted to the BIS MBW outcomes, comparing standard and diffuse prior model-based estimates.

|  | LCI | | CEV (mL) | | FRC (mL) | |
|---|---|---|---|---|---|---|
| **Parameter** | Median | 95% CI | Median | 95% CI | Median | 95% CI |
| ***Mean components*** | | | | | | |
| Standard method ($\alpha_1$) | 6.10 | (6.07, 6.14) | 796.44 | (785.61, 807.47) | 131.00 | (129.1, 132.95) |
| Model-based method: diffuse ($\alpha_2$) | 6.08 | (6.05, 6.12) | 773.11 | (762.43, 783.87) | 127.80 | (125.88, 129.71) |
| *Mean difference ($\alpha_1 - \alpha_2$)* | 0.02 | (0.01, 0.03) | 23.29 | (21.88, 24.69) | 3.20 | (3.02, 3.43) |
| ***Variation components*** | | | | | | |
| Between-participant ($\gamma$) | 0.24 | (0.21, 0.28) | 104.48 | (96.68, 113.02) | 19.20 | (17.86, 20.87) |
| Replicate-participant interaction ($\omega$) | 0.38 | (0.36, 0.4) | 55.08 | (52.29, 58.07) | 9.80 | (9.34, 10.37) |
| Method-participant interaction ($\tau$) | 0.03 | (0, 0.05) | 3.95 | (1.11, 5.73) | 0.90 | (0.68, 1.07) |
| Common residuals ($\sigma$) | 0.14 | (0.137, 0.15) | 15.32 | (14.55, 16.14) | 2.10 | (1.98, 2.18) |

Notes: LCI: lung clearance index; CEV: cumulative expired volume; FRC: functional residual capacity; CI: credible interval.

**Table 4**

**Summary of MCMC convergence for the analysis of shortened tests.** The proportion of tests that exhibited sub-optimal parameter convergence, by having $\hat{R} > 1.25$ for at least one test parameter, in the shortened tests analysis.

|                      | Proportion of tests with convergence issues (%) | |
|----------------------|---------------|-------------------|
| Shortened test length | Diffuse prior | Informative prior |
| 1/3                  | 40            | 18                |
| 1/4                  | 8             | 3                 |
| 1/5                  | <1            | 2                 |
| 1/10                 | <1            | 5                 |
| 1/20                 | <1            | 5                 |
| 1/30                 | <1            | 2                 |
| Complete data        | <1            | 3                 |

# Figures

## Figure 1

**Example of raw MBW data.** Air flow (L/s) and molar mass (g/mol), shown over time (s) for a representative test. The left panels show the data for both the wash-in and washout phases; the right panels show the data for the washout phase only.

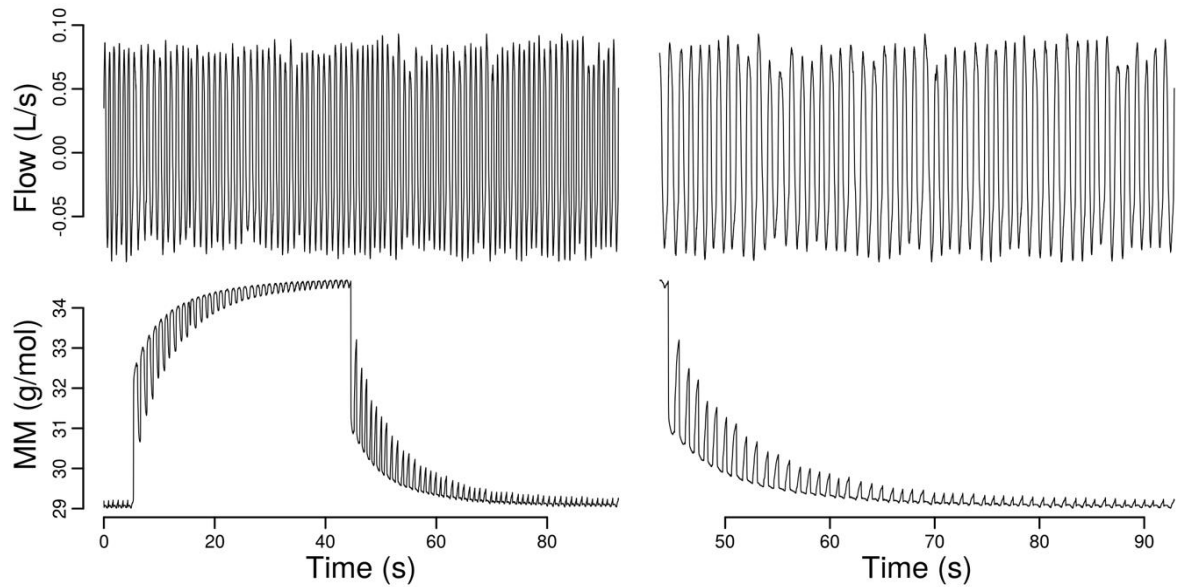

# Figure 2

**Example of derived MBW test quantities.** These are shown for three randomly selected tests. GAS: tracer gas quantity (%); CEVGM: cumulative expired volume of gas mixture (mL); CEVTG: cumulative expired volume of tracer gas (mL).

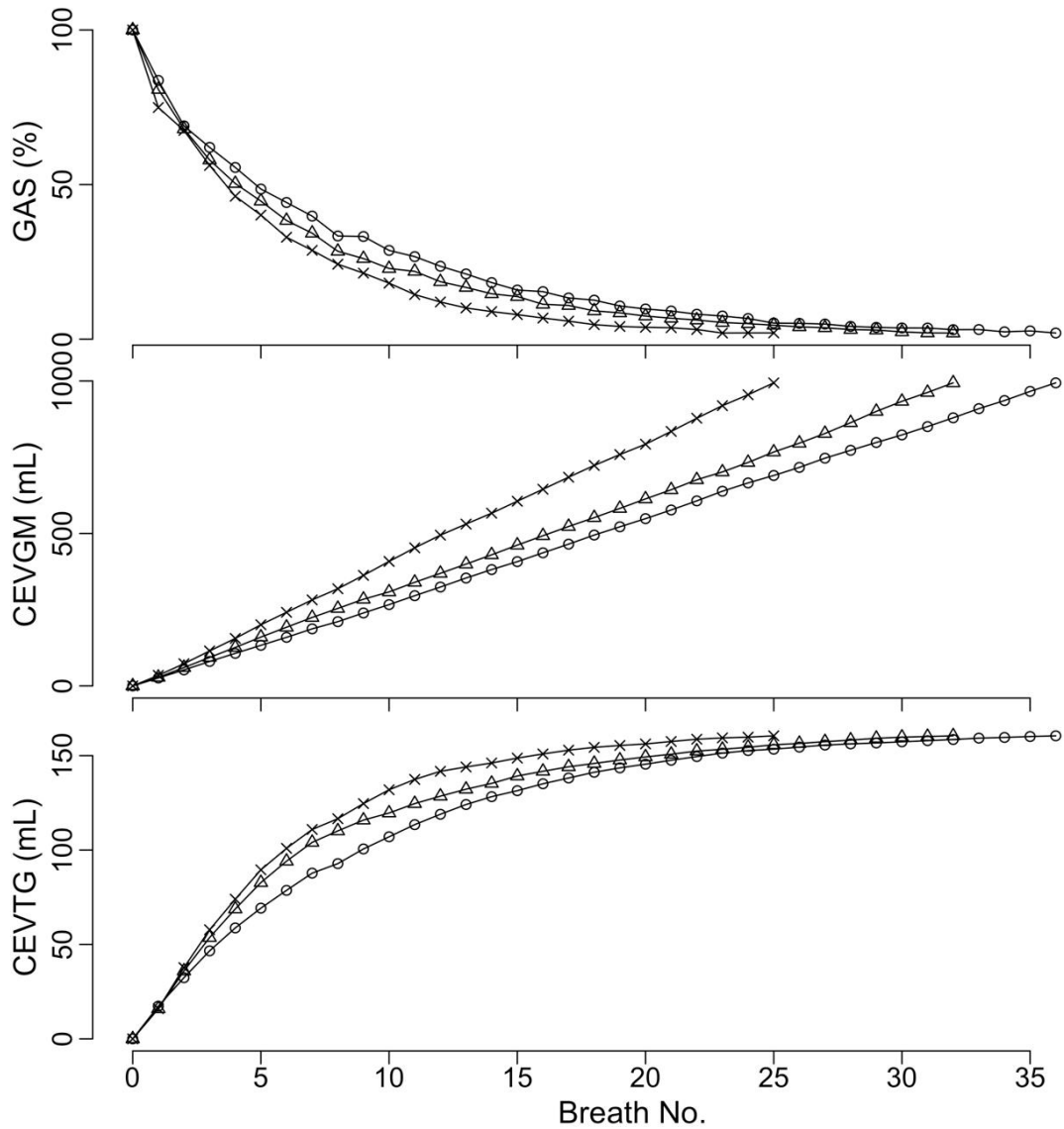

# Figure 3

**Example of a model fit.** Plots showing, for a typical test and informative priors, the observed data (points), model-based mean prediction (black line), and a sample of posterior predictive draws (grey lines). The panels in the left column show the quantities on the transformed scale, those in the right column show them on the original scale.

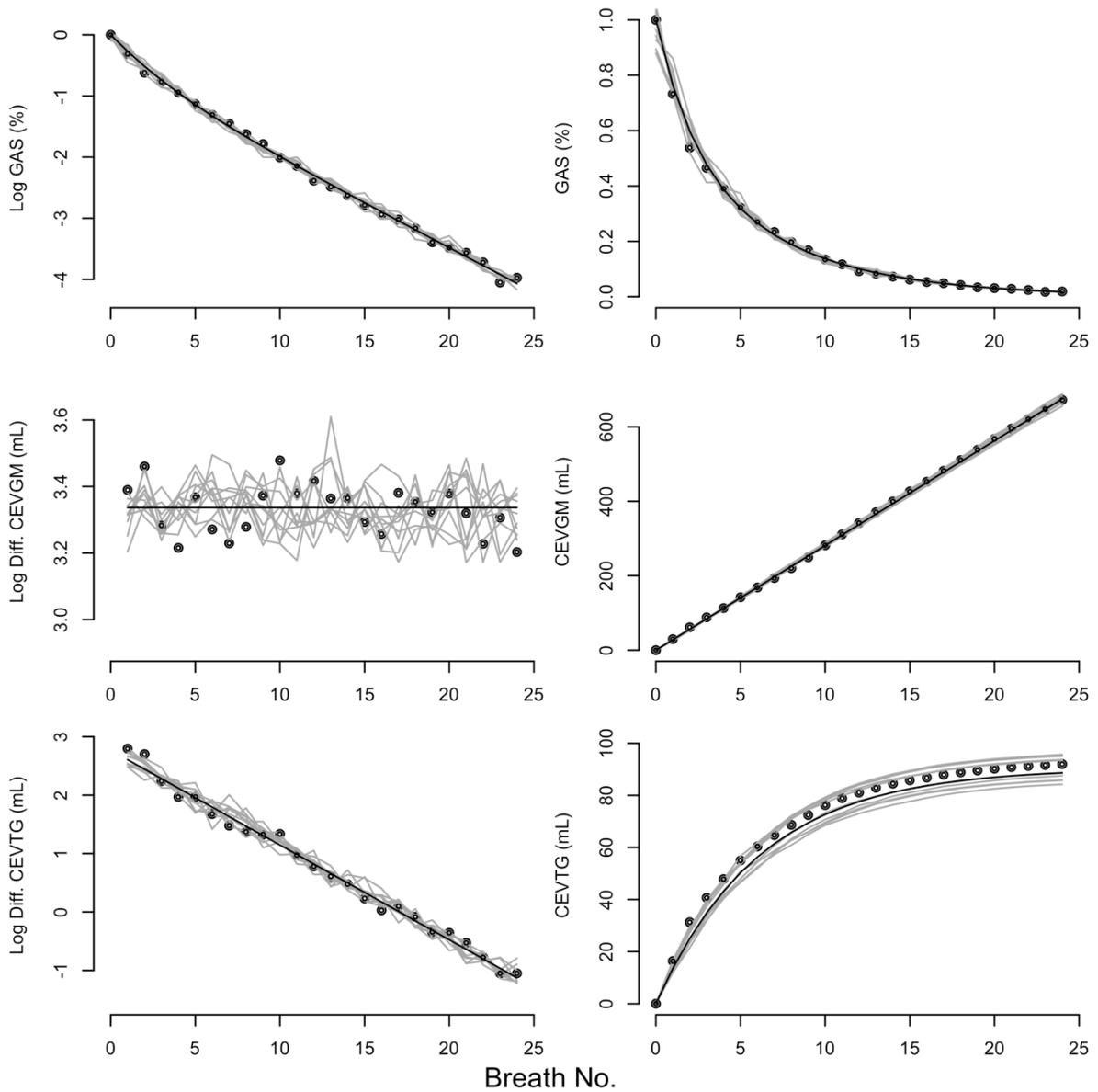

# Figure 4

**Comparison of estimation methods.** Scatterplots showing a comparison of MBW quantities estimated with complete data using the standard method vs model-based method with diffuse ($n = 1{,}197$) and informative priors ($n = 212$). LCI: lung clearance index; CEV: cumulative expired volume (mL); FRC: functional residual capacity (mL).

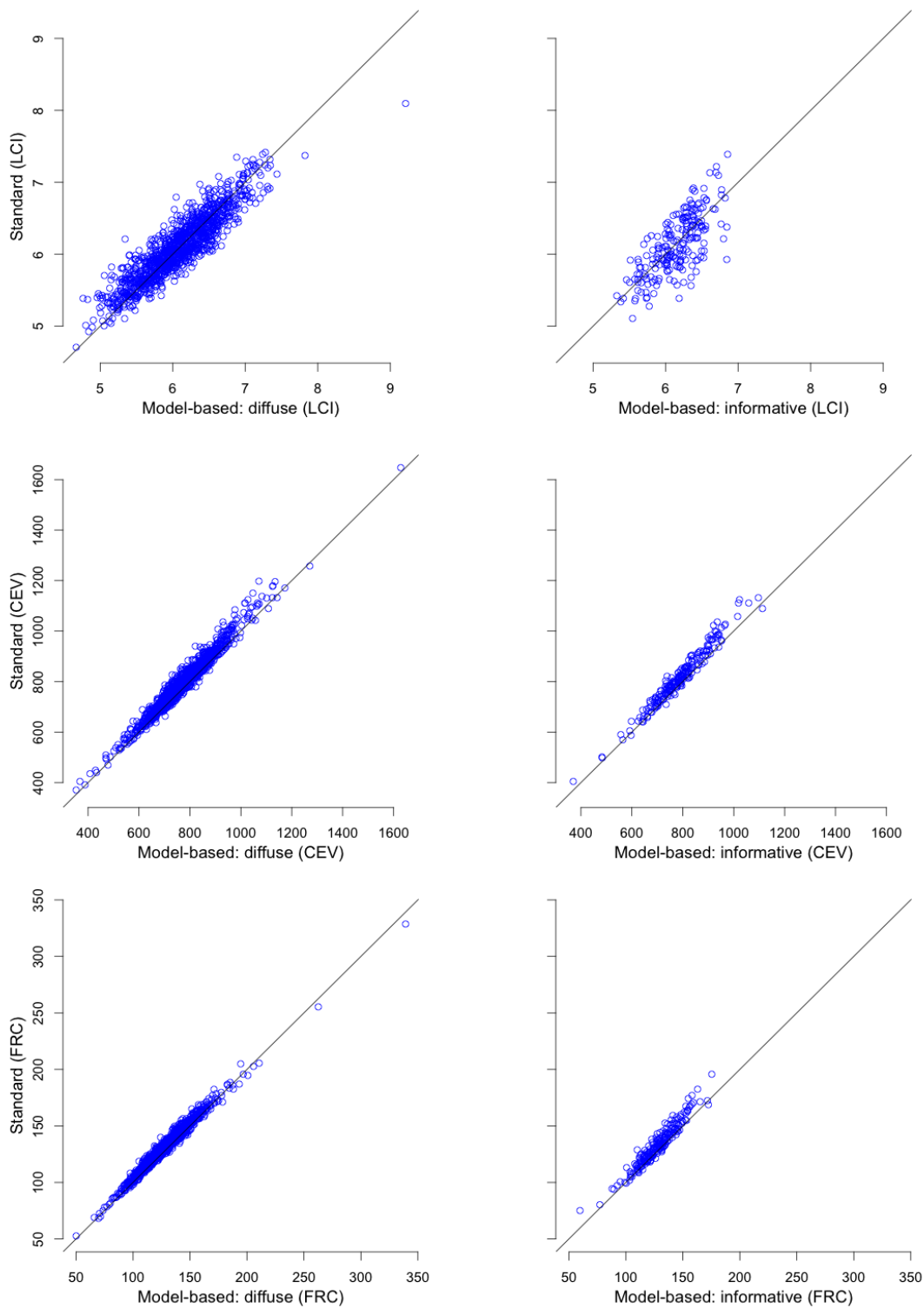

# Figure 5

**Estimation performance for LCI using shortened tests.** The distributions of the posterior median prediction errors (PE) and widths of 95% credible intervals derived from the model-based LCI posteriors, shown for scenarios based on different degrees of test completeness and either the diffuse ($n = 1{,}197$) or informative prior distribution ($n = 212$). Both of the performance measures were first divided by the posterior median using complete test data (on a per-test basis) and are therefore shown as a percentage. Solid blue lines: median of the performance measure across all tests; dashed blue lines: 2.5th and 97.5th percentiles of the performance measure across all tests; shortened test lengths correspond to a tracer gas quantity; 'All': refers to estimates from tests using complete data, included for reference.

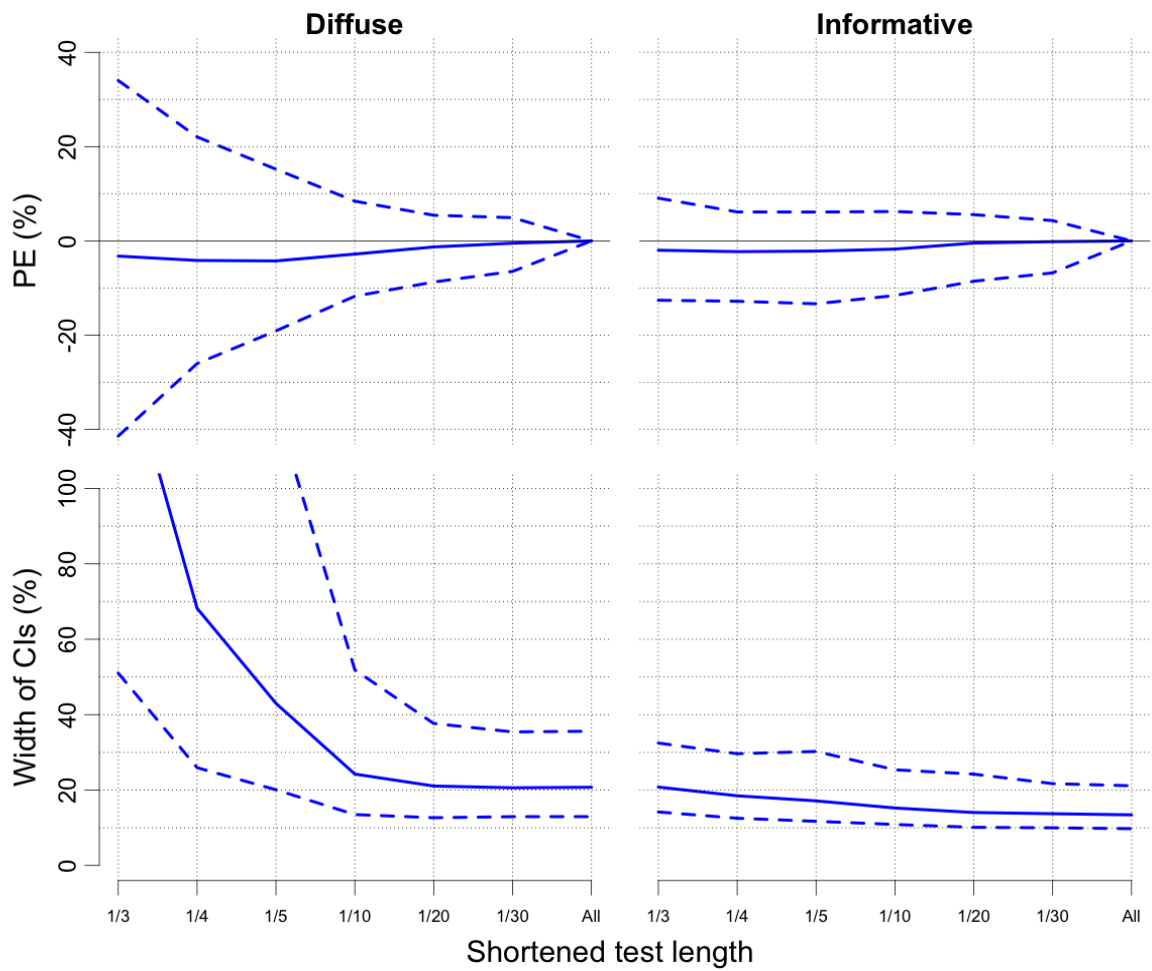

# Figure 6

**Estimation performance for FRC using shortened tests.** The same as Figure 5 but now for estimates of FRC rather than LCI.

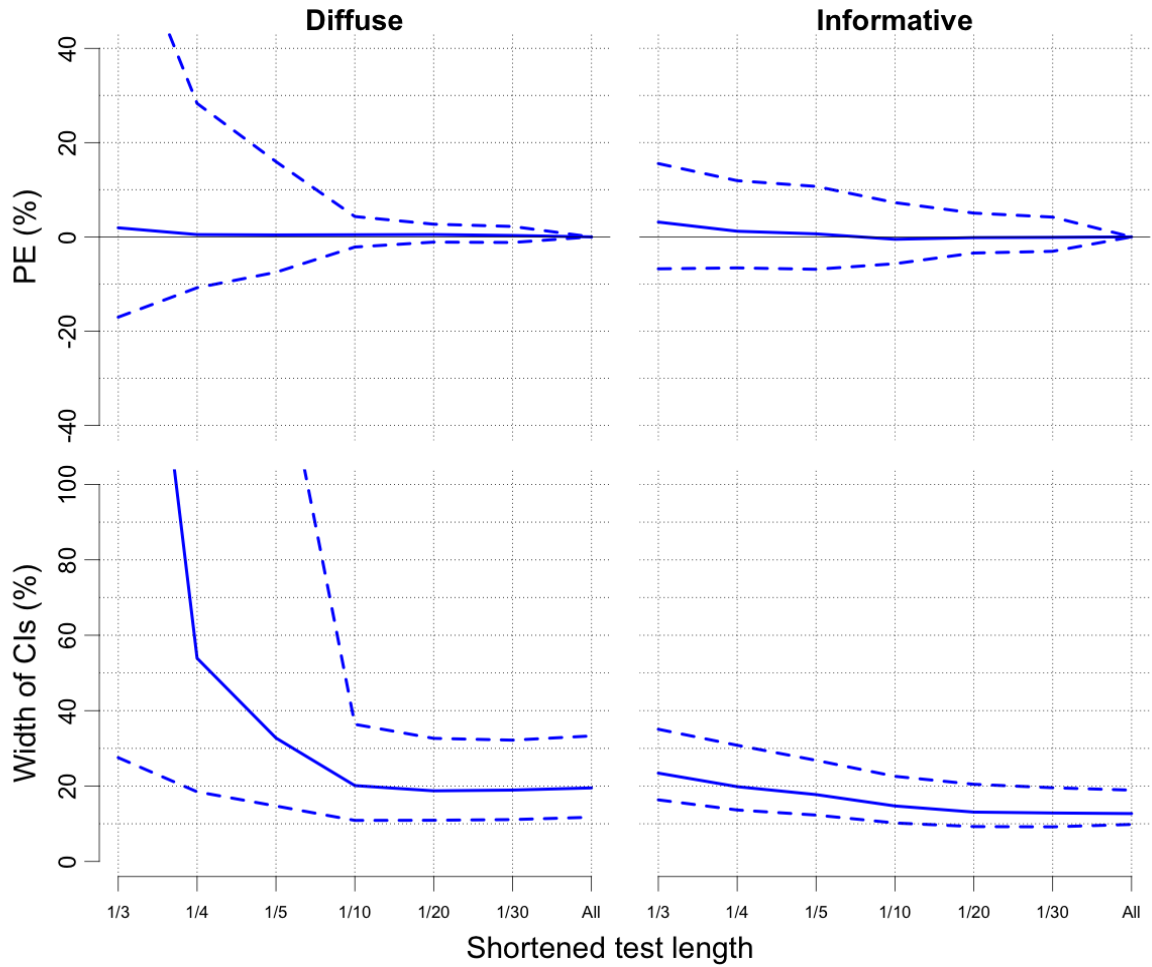

**Figure 7**

**Contrasting alternative model fits on the cumulative scale.** Plots showing, for a typical test, the observed data (points), model-based median prediction (black lines), and a sample of posterior predictive draws (grey lines) for the fit based on (a) a model for the log discrete derivatives, and (b) a model directly for the untransformed cumulative data.

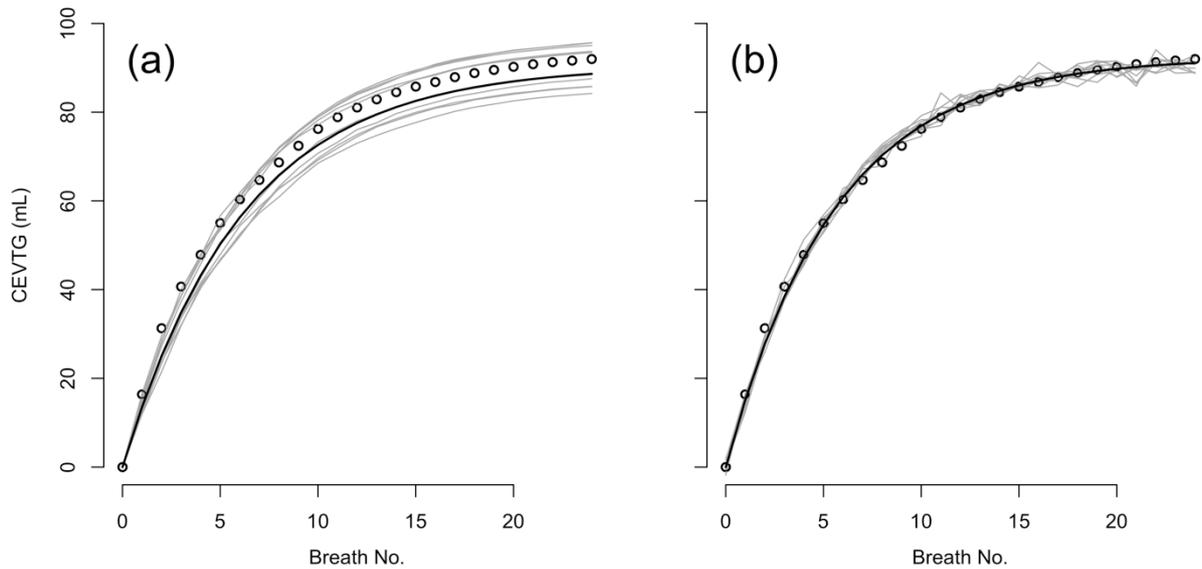

## Figure 8

**Comparison of FRC estimates across replicates.** Estimates of FRC using three different methods: standard method ($FRC^{(s)}$, red circles), model-based method using discrete derivatives ($FRC^{(m^*)}$, black triangles) and a similar model but formulated on the cumulative scale (black squares). The model-based estimates were calculated using the diffuse prior and the resulting 95% credible intervals are shown. Notice that intervals from the model we use (discrete derivatives) more accurately portray the observed variation in FRC estimates across replicates.

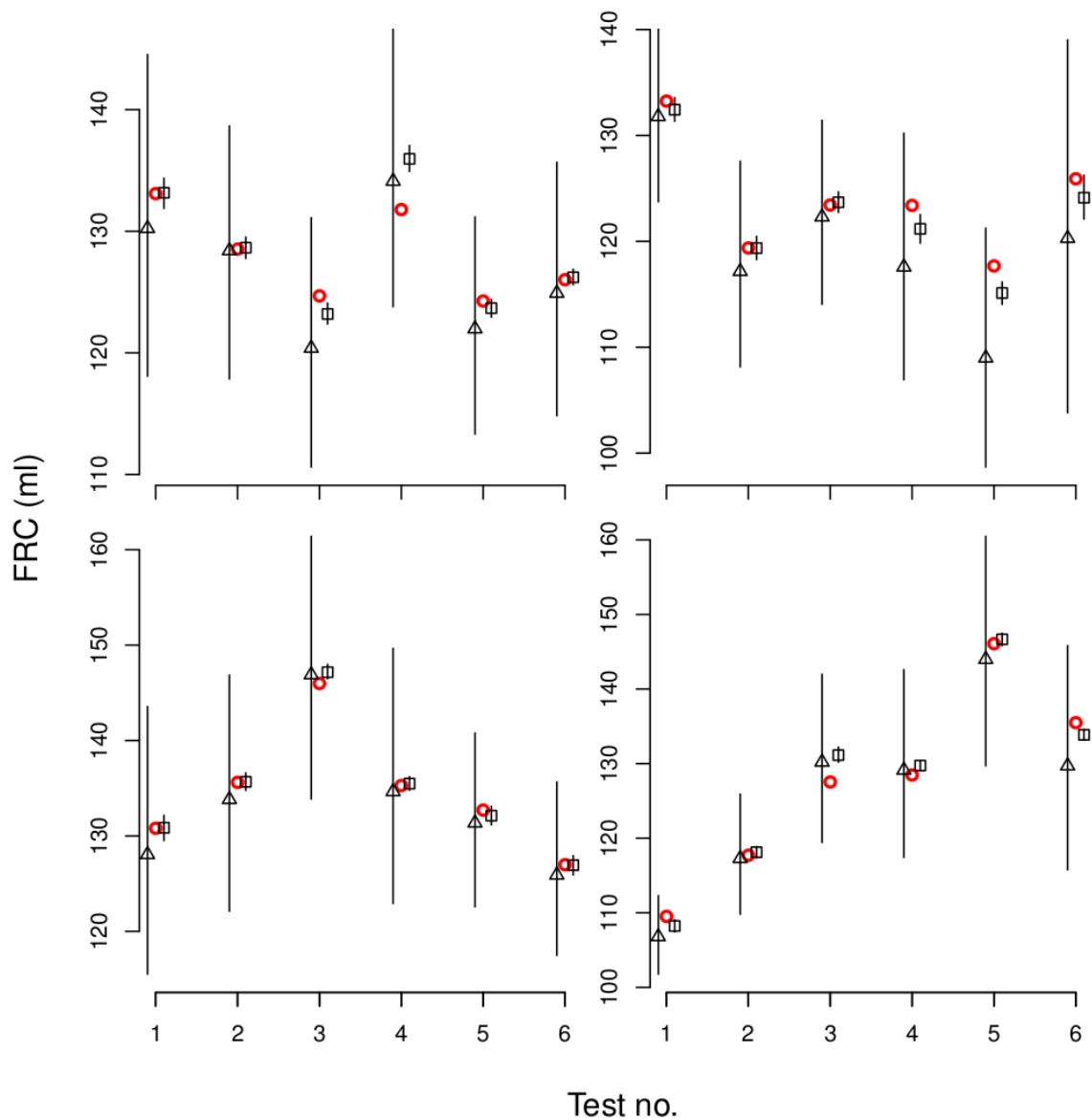

Supplementary material

**Supplementary material for "Bayesian modelling of lung function data from multiple-breath washout tests"**

We include Stan code for four models included in the main manuscript:

- Individual MBW models
    - Using the diffuse prior (`mbw_individual_diffuse.stan`)
    - Using the informative prior (`mbw_individual_informative.stan`)
- Linear mixed-effects (LME) models:
    - Exchangeable replicates and method-specific residuals (`lme_exchangeable_separate.stan`)
    - Linked replicates and common residuals (`lme_linked_common.stan`)

These models require the use of Stan through one of its interfaces, e.g. RStan, PyStan, CmdStan. We used RStan (version 2.12.1) implemented in R (version 3.3.1). More information is available at http://mc-stan.org/.

Data for the individual MBW models (`mbw_individual_diffuse.stan`, `mbw_individual_informative.stan`) should be a list containing the following elements (note that the model using diffuse priors does not require the elements `mu` and `Sigma`):

- `M`: an integer count of the number of washout breaths measured in this test,
  e.g. `int 30`
- `k`: an integer-valued vector containing the index number of each breath (counting from the start of the washout),
  e.g. `num [1:30] 0 1 2 3 4 5 6 7 8 9 ...`
- `gas`: a real-valued vector containing the tracer gas quantity (GAS) at the end of the breath indexed by `k`,
  e.g. `num [1:30] 1 0.701 0.501 0.404 0.390 ...`
- `cevgm`: a real-valued vector containing the observed cumulative expired volume of gas mixture (CEVGM) at the end of the breath indexed by `k`,
  e.g. `num [1:30] 0 36.2 72.0 107.0 141.4 ...`
- `cetgv`: a real-valued vector containing the observed cumulative expired volume of tracer gas (CETGV) at the end of the breath indexed by `k`,
  e.g. `num [1:30] 0 22.1 36.4 46.9 55.0 ...`
- `mu`: a real-valued vector containing 6 elements corresponding to informative prior means for, in order, $\beta_0, \beta_1, \beta_2, \beta_3, \beta_4, \beta_5$.

e.g. `num [1:6] 0.67 0.13 0.14 ...`

- `Sigma`: an real-valued matrix containing the elements of the informative cross-variance matrix corresponding to the values in `mu`.

  e.g. `num [1:6, 1:6]`

  |           | Sigma[,1] | Sigma[,2] | Sigma [,3] | ... |
  |-----------|-----------|-----------|------------|-----|
  | Sigma[1,] | 0.024     | 0.001     | 0.027      | ... |
  | Sigma[2,] | 0.001     | 0.001     | 0.001      | ... |
  | Sigma[3,] | 0.027     | 0.001     | 0.048      | ... |
  | ...       | ...       | ...       | ...        | ... |

Data for the LME models (`lme_exchangeable_separate.stan`, `lme_linked_common.stan`) should be a list containing the following elements (note that the LME model using exchangeable replicates does not require the vector `repl_ind`):

- `n_item`: an integer count of the number of items (participants) in the analysis,

  e.g. `int 10`

- `n_all`: an integer count of the total amount of observations,

  e.g. `int 100`

- `n_meth`: an integer count of the total number of methods to be compared,

  e.g. `int 3`

- `y`: a real-valued vector containing the observations to compare,

  e.g. `num [1:100] 111.1 100.1 120.2 110.8 180.9 ...`

- `item`: an integer-valued vector specifying the item (participant) index for each observation,

  e.g. `int [1:100] 1 1 1 1 1 1 2 2 2 2 ...`

- `meth`: an integer-valued vector specifying the method index for each observation,

  e.g. `int [1:100] 1 1 1 1 1 1 1 1 1 1 ...`

- `repl_ind`: an integer-valued vector specifying the replicate index for each observation (i.e. a separate index for each actual test conducted),

  e.g. `int [1:100] 1 2 3 4 5 6 7 8 9 10 ...`

```stan
functions {
  /**
   * Evaluate the tracer gas quantity at a particular time point.
   * This is a two-component exponential decay model.
   *
   * @param t     Time point at which to evaluate the gas quantity
   * @param beta0 Mixture weight parameter
   * @param beta  Exponents for each component
   *
   * @return Gas quantity (between 0 and 1) at the given time point.
   */
  real gas_curve(real t, real beta0, vector beta) {
    return beta0 * exp(-beta[1] * t) + (1 - beta0) * exp(-beta[2] * t);
  }

  /**
   * Determine if the tracer gas quantity crosses a given threshold
   * between two time points, t1 and t2.
   *
   * @param t1        First time point (left end of desired time
   *                  interval)
   * @param t2        Second time point (right end of desired time
   *                  interval)
   * @param threshold Threshold for the gas quantity
   * @param beta0     Mixture weight parameter
   * @param beta      Exponents for the mixture components
   *
   * @return Boolean value: true if the threshold is crossed, false
   *                  otherwise
   */
  int gas_passes_threshold(real t1, real t2, real threshold, real beta0,
                           vector beta) {
    return gas_curve(t1, beta0, beta) >  threshold &&
           gas_curve(t2, beta0, beta) <= threshold;
  }
}

data {
  int       M ;   // number of observations
  real    gas[M];  // tracer gas quantity (GAS)
  real   cevgm[M];  // cumulative expired volume of gas mixture (CEVGM)
  real   cevtg[M];  // cumulative expired volume of tracer gas  (CEVTG)
  int       k[M];  // breath index
}

transformed data {
  int       nmax        ;
  real   cevgm_tf[M - 1];
  real   cevtg_tf[M - 1];

  // Resolution for line search used to deterimine the real-valued
  // stopping 'breath'.
  nmax = 100;

  // Transform cumulative data to log increment scale.
  for (m in 1:(M - 1)) {
    cevgm_tf[m] = log(cevgm[m + 1] - cevgm[m]);
    cevtg_tf[m] = log(cevtg[m + 1] - cevtg[m]);
```

```stan
  }
}

parameters {
  // GAS curve.
  real<lower=0, upper=1>  beta0;
  positive_ordered[2]     beta;  // a.k.a 'beta1' and 'beta2' (log scale)
  real<lower=0>           error_gas;

  // CEVGM curve.
  real<lower=0>           beta5;
  real<lower=0>           error_cevgm;

  // CEVTG curve.
  real<lower=0>           beta3;
  real<lower=0>           beta4;
  real<lower=0>           error_cevtg;
}

transformed parameters {
  // Stopping breaths.
  real<lower=0>  m_stop_40;
  real<lower=0>  m_stop_40_cont;

  // MBW quantities using standard stopping breath.
  real<lower=0>  lci40;
  real<lower=0>  cev40;

  // MBW quantities using model-based stopping breath.
  real<lower=0>  lci40_cont;
  real<lower=0>  cev40_cont;

  // Define MBW quantities from model parameters.
  // Use a threshold of 1/40 (= 0.025) for the gas concentration.
  for (m in 2:100) {
    if (gas_passes_threshold(m - 1, m, 0.025, beta0, beta)) {
      // Use the standard stopping breath, k(40).
      m_stop_40 = m;
      cev40 = beta5 * m_stop_40;
      lci40 = cev40 / beta3;   // n.b. FRC = beta3

      // Use the model-based stopping 'breath', theta.
      // This stopping breath is real-valued rather than integer-valued.
      // It is determined by a simple line search.
      for (i in ((m - 1) * nmax):(m * nmax)) {
        if (gas_passes_threshold(1.0 / nmax * (i - 1),
                                 1.0 / nmax *  i,  0.025, beta0, beta)) {
          m_stop_40_cont = 1.0 / nmax * i;
          cev40_cont = beta5 * m_stop_40_cont;
          lci40_cont = cev40_cont / beta3;   // n.b. FRC = beta3
        }
      }
    }
  }
}
```

```
  model {
    // Declare local variables.
    // These are used to store the mean and sd of the 3 curves, see below.
    vector[M]     gas_mean;
    vector[M - 1] cevgm_mean;
    vector[M - 1] cevtg_mean;
    vector[M]     gas_sd;
    vector[M - 1] cevgm_sd;
    vector[M - 1] cevtg_sd;

    // Priors (diffuse).

    beta0   ~ beta(2, 2);   // a.k.a. beta0
    beta[1] ~ normal(0, 1);
    beta[2] ~ normal(0, 1);

    beta3 ~ normal(0, 1000);
    beta4 ~ normal(0, 1);

    beta5 ~ normal(0, 100);

    error_gas   ~ cauchy(0, 2.5);
    error_cevgm ~ cauchy(0, 2.5);
    error_cevtg ~ cauchy(0, 2.5);

    // Calculate the mean and sd of the 3 curves.
    for (m in 1:M) {
      gas_mean[m] = gas_curve(k[m], beta0, beta);
      gas_sd[m]   = error_gas;
    }
    for (m in 1:(M - 1)) {
      cevgm_mean[m] = log(beta5);
      cevtg_mean[m] = log(beta3) + log(1 - exp(-beta4)) - (beta4 * (k[m]));

      cevgm_sd[m] = error_cevgm;
      cevtg_sd[m] = error_cevtg;
    }

    // Define the probability models for the 3 curves.
    gas      ~ lognormal(log(gas_mean),  gas_sd);
    cevgm_tf ~    normal(   cevgm_mean , cevgm_sd);
    cevtg_tf ~    normal(   cevtg_mean , cevtg_sd);
  }
```

```
functions {
  /**
   * Evaluate the tracer gas quantity at a particular time point.
   * This is a two-component exponential decay model.
   *
   * @param t     Time point at which to evaluate the gas quantity
   * @param beta0 Mixture weight parameter (a.k.a. beta0)
   * @param beta  Exponents for each component
   *
   * @return Gas quantity (between 0 and 1) at the given time point.
   */
  real gas_curve(real t, real beta0, vector beta) {
    return beta0 * exp(-beta[1] * t) + (1 - beta0) * exp(-beta[2] * t);
  }

  /**
   * Determine if the tracer gas quantity crosses a given threshold between
   * two time points, t1 and t2.
   *
   * @param t1        First time point (left end of desired time
   *                  interval)
   * @param t2        Second time point (right end of desired time
   *                  interval)
   * @param threshold Threshold for the gas quantity
   * @param beta0     Mixture weight parameter (a.k.a. beta0)
   * @param beta      Exponents for the mixture components
   *
   * @return Boolean value: true if the threshold is crossed, false
   *                        otherwise
   */
  int gas_passes_threshold(real t1, real t2, real threshold, real beta0,
                           vector beta) {
    return gas_curve(t1, beta0, beta) >  threshold &&
           gas_curve(t2, beta0, beta) <= threshold;
  }
}

data {
  int       M ;   // number of observations
  real   gas[M];  // tracer gas quantity (GAS)
  real   cevgm[M]; // cumulative expired volume of gas mixture (CEVGM)
  real   cevtg[M]; // cumulative expired volume of tracer gas  (CEVTG)
  int       k[M]; // breath index
  vector[6]       mu; // informative prior means
  matrix[6, 6] Sigma; // informative cross-covariance matrix
}

transformed data {
  int       nmax        ;
  real   cevgm_tf[M - 1];
  real   cevtg_tf[M - 1];

  // Resolution for line search used to deterimine the real-valued
  // stopping 'breath'.
  nmax = 100;
```

```
  // Transform cumulative data to log increment scale.
  for (m in 1:(M - 1)) {
    cevgm_tf[m] = log(cevgm[m + 1] - cevgm[m]);
    cevtg_tf[m] = log(cevtg[m + 1] - cevtg[m]);
  }
}

parameters {
  // GAS curve.
  real<lower=0, upper=1>  beta0;
  positive_ordered[2]     beta; // a.k.a 'beta1' and 'beta2'
  real<lower=0>      error_gas;

  // CEVGM curve.
  real<lower=0>         beta5;
  real<lower=0>   error_cevgm;

  // CEVTG curve.
  real<lower=0>         beta3;
  real<lower=0>         beta4;
  real<lower=0>   error_cevtg;
}

transformed parameters {

  // Vector to keep different parameters in when specifying multi_normal
  // prior.
  vector[6] zeta;

  // Stopping breaths.
  real<lower=0>  m_stop_40;
  real<lower=0>  m_stop_40_cont;

  // MBW quantities using standard stopping breath.
  real<lower=0>   lci40;
  real<lower=0>   cev40;

  // MBW quantities using model-based stopping breath.
  real<lower=0>   lci40_cont;
  real<lower=0>   cev40_cont;

  // Store parameters in single vector.
  zeta[1] = beta0;
  zeta[2] = beta[1];
  zeta[3] = beta[2];
  zeta[4] = beta3;
  zeta[5] = beta4;
  zeta[6] = beta5;

  // Define MBW quantities from model parameters.
  // Use a threshold of 1/40 (= 0.025) for the gas concentration.
  for (m in 2:100) {
    if (gas_passes_threshold(m - 1, m, 0.025, beta0, beta)) {
      // Use the standard stopping breath, k(40).
      m_stop_40 = m;
      cev40 = beta5 * m_stop_40;
      lci40 = cev40 / beta3;   // n.b. FRC = beta3
```

```
      // Use the model-based stopping 'breath', theta.
      // This stopping breath is real-valued rather than integer-valued.
      // It is determined by a simple line search.
      for (i in ((m - 1) * nmax):(m * nmax)) {
        if (gas_passes_threshold(1.0 / nmax * (i - 1),
                                 1.0 / nmax *  i,  0.025, beta0, beta)) {
          m_stop_40_cont = 1.0 / nmax * i;
          cev40_cont = beta5 * m_stop_40_cont;
          lci40_cont = cev40_cont / beta3;  // n.b. FRC = beta3
        }
      }
    }
  }
}

model {
  // Declare local variables.
  // These are used to store the mean and sd of the 3 curves, see below.
  vector[M]     gas_mean;
  vector[M - 1] cevgm_mean;
  vector[M - 1] cevtg_mean;
  vector[M]     gas_sd;
  vector[M - 1] cevgm_sd;
  vector[M - 1] cevtg_sd;

  // Informative multivariate normal prior.

  zeta ~ multi_normal(mu, Sigma);
  error_gas   ~ cauchy(0, 2.5);
  error_cevtg ~ cauchy(0, 2.5);
  error_cevgm ~ cauchy(0, 2.5);

  // Calculate the mean and sd of the 3 curves.
  for (m in 1:M) {
    gas_mean[m] = gas_curve(k[m], beta0, beta);
    gas_sd[m]   = error_gas;
  }
  for (m in 1:(M - 1)) {
    cevgm_mean[m] = log(beta5);
    cevtg_mean[m] = log(beta3) + log(1 - exp(-beta4)) - (beta4 * (k[m]));

    cevgm_sd[m] = error_cevgm;
    cevtg_sd[m] = error_cevtg;
  }

  // Define the probability models for the 3 curves.
  gas      ~ lognormal(log(gas_mean),   gas_sd);
  cevgm_tf ~    normal(   cevgm_mean , cevgm_sd);
  cevtg_tf ~    normal(   cevtg_mean , cevtg_sd);
}
```

```
data {
  int<lower=0>       n_all ;  // number of observations
  int<lower=0>       n_item;  // number of participants (items)
  int<lower=0>       n_meth;  // number of methods
  real               y[n_all];  // outcome variable
  int<lower=1>  item[n_all];   // item index
  int<lower=1>  meth[n_all];   // method index
}

transformed data {
  int<lower=1>  n_factor;
  n_factor = n_all / n_meth;
}

parameters {
  // Parameters for each method.
  vector[n_meth]  alpha;  // mean effect
  vector[n_meth]  sigma;  // sd

  // Random effects due to item.
  vector[n_item]  u_raw;  // per-item random effect
  real<lower=0>   gamma;  // sd

  // Random effects due to interaction between items and methods.
  vector[n_item]  c_raw[n_meth];  // per-combination random effect
  real<lower=0>     tau       ;  // sd
}

model {
  vector[n_item]  u         ;  // per-item random effect
  vector[n_item]  c[n_meth];  // per-combination random effect

  // Prior for fixed components.
  alpha ~ normal(0, 10000);

  // Priors for variance components.
  gamma ~ cauchy(0, 2.5);
  sigma ~ cauchy(0, 2.5);
  tau   ~ cauchy(0, 2.5);

  // Distributions of random effects.
  u_raw ~ normal(0, 1);
  u = gamma * u_raw;

  for (i in 1:n_meth) {
    c_raw[i] ~ normal(0, 1);
    c[i] = tau * c_raw[i];
  }

  // Distribution of the outcome variable.
  for (n in 1:n_all) {
    y[n] ~ normal(alpha[meth[n]] +
                  u[item[n]] +
              c[meth[n], item[n]],
                sigma[meth[n]]);
  }
}
```

lme_exchangeable_separate.stan

```
generated quantities {
  real<lower=0> sigma_ratio[1];
  real alpha_diff[1];

  sigma_ratio[1] <- sigma[1]/sigma[2];
  alpha_diff[1]  <- alpha[1]-alpha[2];
}
```

# lme_linked_common.stan

```stan
data {
  int<lower=0>             n_all ;  // number of observations
  int<lower=0>             n_item;  // number of participants (items)
  int<lower=0>             n_meth;  // number of methods
  real                     y[n_all];  // outcome variable
  int<lower=1>      item[n_all];  // item index
  int<lower=1>      meth[n_all];  // method index
  int<lower=1>  repl_ind[n_all];  // item-replicate index
}

transformed data {
  int<lower=1>  n_factor;
  n_factor = n_all / n_meth;
}

parameters {
  // Parameters for each method.
  vector[n_meth]  alpha;  // mean effect
  real            sigma;  // sd

  // Random effects due to item.
  vector[n_item]  u_raw;  // per-item random effect
  real<lower=0>   gamma;  // sd

  // Random effects due to interaction between items and methods.
  vector[n_item]  c_raw[n_meth];  // per-combination random effect
  real<lower=0>    tau          ;  // sd

  // Random effects due to interaction between items and replicates.
  vector[n_factor]  a_raw;  // item-replicate interaction effect
  real<lower=0>     omega;  // sd
}

model {
  vector[n_item]    u         ;  // per-item random effect
  vector[n_item]    c[n_meth];  // per-combination random effect
  vector[n_factor]  a         ;  // item-replicate interaction effect

  // Prior for fixed components.
  alpha ~ normal(0, 10000);

  // Priors for variance components.
  gamma ~ cauchy(0, 2.5);
  sigma ~ cauchy(0, 2.5);
  tau   ~ cauchy(0, 2.5);
  omega ~ cauchy(0, 2.5);

  // Distributions of random effects.
  u_raw ~ normal(0, 1);
  u = gamma * u_raw;

  for (i in 1:n_meth) {
    c_raw[i] ~ normal(0, 1);
    c[i] = tau * c_raw[i];
  }
```

```
  for (i in 1:n_factor) {
    a_raw[i] ~ normal(0, 1);
    a[i] = omega * a_raw[i];
  }
  
  // Distribution of the outcome variable.
  for (n in 1:n_all) {
    y[n] ~ normal(alpha[meth[n]] +
                    u[item[n]] +
                  a[repl_ind[n]] +
            c[meth[n], item[n]],
                          sigma);
  }
}

generated quantities {
  real alpha_diff[1];
  alpha_diff[1]  <- alpha[1]-alpha[2];
}
```